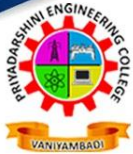

# Priyadarshini Engineering College

(Approved by AICTE, New Delhi and Permanently Affiliated to Anna University, Chennai)
Chettiyappanur Village & Post, Vaniyambadi-635751, Vellore District, Tamil Nadu, India.
Listed in 2(f) & 12(B) Sections of UGC.

Technical Sponsor by

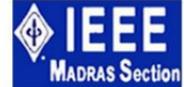

## CERTIFICATE

### International Conference on Electrical, Electronics, Computers, Communication, Mechanical and Computing (EECCMC) - 2018

This is to certify that **Mr. Seyyedmostafa Mousavi Janbehsarayi** has presented a paper entitled: **An Analytical Solution for Nonlinear Dynamics of One Kind of Scanning Probe Microscopes** with paper code: **01-2018-400** in International Conference on Electrical, Electronics, Computers, Communication, Mechanical and Computing (EECCMC) - 2018, with catalog "CFP18O37-PRT: 978-1-5386-4303-7", organized by Priyadarshini Engineering College, Vellore District, Tamil Nadu, India during 28th & 29th January 2018.

*IEEE Conference Record # 43456*

Certificate Proof

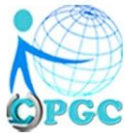

Dr. Siva Ganesh Malla
Director, CPGC

Dr. P. Natarajan
Principal, PEC

# CERTIFICATE

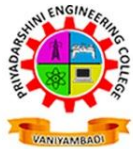

## Priyadarshini Engineering College

(Approved by AICTE, New Delhi and Permanently Affiliated to Anna University, Chennai)
Chettiyappanur Village & Post, Vaniyambadi-635751, Vellore District, Tamil Nadu, India.

Listed in 2(f) & 12(B) Sections of UGC.

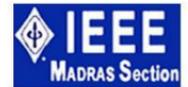

Technical Sponsor by
IEEE MADRAS Section

### International Conference on Electrical, Electronics, Computers, Communication, Mechanical and Computing (EECCMC) - 2018

This is to certify that **Mr. Seyyedmostafa Mousavi Janbehsarayi** has published a paper entitled: **An Analytical Solution for Nonlinear Dynamics of One Kind of Scanning Probe Microscopes** with paper code: **01-2018-400** in International Conference on Electrical, Electronics, Computers, Communication, Mechanical and Computing (EECCMC) - 2018, with catalog "CFP18O37-PRT: 978-1-5386-4303-7", organized by Priyadarshini Engineering College, Vellore District, Tamil Nadu, India during 28th & 29th January 2018.

Certificate Proof

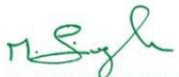

Dr. Siva Ganesh Malla
Director, CPGC

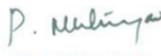

Dr. P. Natarajan
Principal, PEC

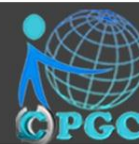

PGC

# An analytical solution for nonlinear dynamics of one kind of scanning probe microscopes

Sina Eftekhar

*Department of Mechanical Engineering, College of Engineering, University of Tehran, Tehran, Iran*

SeyyedMostafa Mousavi JanbehSarayi

*Department of Mechanical Engineering, College of Engineering, University of Tehran, Tehran, Iran*

*Abstract*— The dynamic behavior of AFM is studied taking into account the nonlinear interaction forces between probe and sample. The exerted forces on the free end of micro-beam are simulated with the third degree polynomial. The effect of some parameters on the dynamics of AFM is studied. The results show that the frequency response of AFM is not sensitive to the tip mass in the case of both the sample and cantilever vibration. The effect of sample vibration is studied in the case of in-phase and anti-phase vibration. The results show that although the vibration amplitude of sample is very small compared to the amplitude of cantilever, it has great effect on the resonant frequency of the cantilever.

*Keywords*

*Nonlinear vibration; tip-mass; resonant frequency*

1. **Introduction:**

The atomic force microscope (AFM) is developed in such a way that can be utilized to measure the intermolecular forces with atomic resolution which can be used in variety of applications like biology[1-2], polymer[3-4], electronics, and materials science. Contact, Non-contact, and tapping mode are three main techniques which can be used for sample topography depending on the properties of the sample. In contact mode, the probe is in constant contact with sample while the interaction forces between probe and sample and the cantilever deflection reveal the topography of the surface[5]. Although this method is very useful for scanning the hard materials, it can't be used for soft materials due to surface damage. For scanning the soft materials, it is recommended to use non-contact (NC) or tapping mode (TM). In these two modes, the cantilever vibrates at a frequency close to its primary resonant frequency[6] while interaction forces affect the amplitude and the phase of vibration.

Most of the failures in topographical image with high resolution rely in setting the working frequency of micro-cantilever close to its resonant frequencies. Many attempts have been devoted to find the effect of cantilever parameters like tip-mass, tilting angle[7], length, width of cantilever, and tip-mass ratio on resonant frequency [8]. In this matter, W.chang et al [9] studied the effect of contact stiffness on the dynamics of AFM using the linear visco-elastic forces for the side-wall probe. In earlier studies of AFM, the interaction forces between probe and sample were simulated with linear visco-elastic forces [9-12]. But as it is expected, the results were far from reality. Rather than reliability of the results, many physical phenomenons like bifurcation and jump cannot be studied in linear systems. K.Yagasaki [13] studied the bifurcation and chaotic response of micro-beam and their effects using the averaging method. They showed that abundant bifurcation and chaotic behavior can occur during the scanning. The interaction forces between probe and sample are strongly nonlinear which make it difficult to model, analyze, and study the dynamics of AFM. There are numerous models of tip-sample interaction forces like Hertz contact model [14], Piecewise linear contact model [15], Derjaguin-Muller-Toporov (DMT) [16], DerjaguinLandau-Verwey-Overbeek (DLVO) theory, a combination of the van der Waals attraction and the electrostatic repulsion between two surfaces in a liquid environment [17], Chadwick for the thin membranes [18], or Kelvin-Voigt dissipation model [19].

Rather than interaction forces which must be modeled in order to study the dynamics of AFM, due to the complicity of solutions, the micro-cantilever, and the governing equations of the motion are needed to be simplified too. Lumped mass model and Euler-Bernoulli beam theory are two different methods for simulating the micro-beam, which have been utilized in variety of papers so far [8,20]. In lumped mass theory, the whole cantilever considered as a mass which represents the cantilever's mass along with springs and dampers in 3-D directions which represent the interaction forces between probe and sample. Inui [21] studied the effect of nonlinear interaction forces on the amplitude and phase of the cantilever using perturbation method and lumped-mass simulation for S-systems (The system in which the sample vibrates and the cantilever is at rest) and P-systems (The system in which cantilever vibrates by external forces and the sample is at rest). He found that both the vibrating amplitude and frequency shift of cantilever is dependant of the force gradient even if the cantilever is not forced to vibrate. The combination of P-systems and S-system in which both sample and cantilever vibrate couldn't be analyzed by his method due to complicity. H.Pishkenari et al [22] studied the deterministic and random excitation of the micro-beam using two degree of freedom for the lumped-mass cantilever and the nonlinear interaction forces. They found the parameter region where the chaotic motion occurs. However, because the lumped-mass model lacks the properties of micro-cantilevers, the results are not reliable, especially when the cantilever's length is not small. Euler-Bernoulli is another theory which has been used in many cases. In this model, the micro-cantilever is replaced by a beam and interaction forces exerted on the free end of micro-cantilever. If the shear deformation and rotary inertia effects become important, Timoshenko's beam assumption can be used to study the dynamics of AFM. In this matter, H.Lee et al [23] used the effect of thermal vibration of AFM on the resonant frequency of micro-beam using continues Timoshenko beam assumption. S.Eslami et al [24] used Timoshenko beam theory to present a comprehensive model for topography of buried materials. Using the tuning mechanical dynamics of AFM system, they proposed a method to reveal the properties of buried materials which can be utilized to imagine the internal composition of cells.

Despite all simulations and modeling of micro-cantilever and interaction forces, for solving the equation of the motion and boundary conditions of a micro-cantilever in nonlinear case, more simplifications, and assumptions are needed. This is because the nonlinear equations and boundary condition problems in nonlinear cases have not the routine solutions. There are several approximate methods which can be utilized to solve the nonlinear vibration of structures like epsilon method and Galerkin's approximation. Using epsilon method needs more precaution, because the reliability of the results strongly depends on the order of epsilon. In all of these cases, the correctness of the results is very limited. The effect of lateral forces must be simplified too. On the other hand, considering the lateral forces makes the understanding and studying the dynamics of AFM much difficult, especially in nonlinear interaction forces. It has been shown that undesired lateral motions have the prominent effect on the vibration of micro-cantilever in normal direction [8]. Finally, when the probe-sample separation become small, the sample surface moves toward and backward the micro-cantilever because of strong electrostatic forces between probe and sample. Also, in some techniques of AFM topography, the sample needs to vibrate too. This method can be utilized in order to reveal the properties of buried materials. It has been shown [21] that both frequency shift and vibrating amplitude is proportional to the vibrating sample. Although the amplitude of vibrating sample is very small compared with the amplitude of cantilever, the vibration of the sample cannot be neglected..

In this paper, the effects of interaction forces in VE (vertical excitation) mode are studied taking into account the nonlinear interaction forces between probe and sample. Interaction forces between probe and sample is simulated with a polynomial. The presented model is in good agreement with exact value and the error of solving the boundary condition problems are abolished by finding the analytical response of the system. The effects of cantilever properties like tip-mass ratio and the amplitude of the vibrating sample in VE mode are obtained.

## 2. Modeling the interaction forces:

The interaction forces between probe and sample are simulated with interactions between sphere and flat surface and considering the long range attractive forces as van Der Waals force [25]. The short range repulsive forces depend on several factors like stiffness, hardness, and tip-ratio. For soft materials with high adhesion forces, the interaction force between tip and sample (DMT model) is:

$$f_n = \begin{cases} -\dfrac{HR}{6d_n^2}, d_n \rangle a_0 \\ -\dfrac{HR}{6a_0^2} + \dfrac{4}{3}E^*\sqrt{R}(a_0-d_n)^{3/2} - \eta_n(a_0-d_n)^{1/2}, d_n \le a_0 \end{cases} \quad 1$$

H, R, $d_n$, $a_0$, $E^*$, $\eta_n$ are Hamaker constant, tip radius, transient tip-sample separation given by $d_n = D + \Delta_n - a_n$ where D is the equilibrium tip-sample separation, $a_n$ is the normal displacement of the sample surface, intermolecular distance, effective elastic modulus given by $\dfrac{1}{E^*} = \dfrac{(1-v_t^2)}{E_t} + \dfrac{(1-v_s^2)}{E_s}$ and viscosity of tip-sample contact in the normal direction respectively. Table.1 shows constants and properties of the micro-cantilever which is used in this paper. When the distance between probe and sample increases, the predominate forces change from repulsive electrostatic forces to attractive van der waals. The linear forces are in good agreement for repulsive regime but when the separation between probe and sample increases, this linearization is not reliable. A good approximation for the nonlinear forces must be done This approximation is in well agreement for both repulsive and attractive regimes for this interval. The maximum error between the exact value of interaction forces and nonlinear approximation used in this paper for both van der waals and electrostatic regime is 7%. We used a polynomial for nonlinear part while interactions between probe and sample are modeled with linear forces.

Table.1: Constants and properties of the micro-cantilever

| Property | Value |
| --- | --- |
| Hamaker constant | $2.96 \times 10^{-11} j$ |
| Tip Radius | 10 nm |
| Cantilever length | 90 μm |
| Cantilever width | 35 μm |
| Cantilever thickness | 1.84 μm |
| Cantilever cross section area | $6.57 \times 10^{-11} m^2$ |
| Cantilever material density | $2300 \, kg/m^3$ |
| Cantilever Young's Modulus | 167 GPa |
| Effective elastic Modulus | 10.7 GPa |
| Intermolecular distance | 0.1923 nm |

### 2.1. Governing equation of the motion and boundary conditions

By considering the linear vibration for a beam with clamped-free ends, and with respect to coupled motion of micro-cantilever, the governing equation of the motion and the boundary conditions for a micro-cantilever with external interaction forces are:

$$\rho A (\frac{\partial^2 y(x,t)}{\partial t^2}) + EI \frac{\partial^4 y(x,t)}{\partial x^4} = 0$$

$$y(0,t) = y_0$$

$$y'(0,t) = 0 \qquad \qquad 2$$

$$EIy''(L,t) = 0$$

$$EIy'''(L,t) = k_1[y(L) - \varepsilon a_n] + k_n[y(L) - \varepsilon a_n]^3 + m_t \ddot{y}$$

$\rho, A, E, I, y(x,t), k_1, k_n, m_t$ are density of the cantilever, cross-section area of cantilever, area moment of inertia around z-axis, displacement of the cantilever in the normal direction, linear spring, nonlinear spring (which is used for normal nonlinear forces) and the tip-mass respectively. Fig.1 shows a cantilever and its properties. As it is mentioned before, the displacement of sample is very small compared with the amplitude of cantilever, so we used $\varepsilon$ in Eq.2. The first equation is the equation of the motion and four other equations are boundary conditions. The general solution of equation 2.a is:

$$y(x,t) = \{a_1 \cosh(\lambda x) + a_2 \cos(\lambda x) + a_3 \sinh(\lambda x) + a_4 \sin(\lambda x)\} \sin(\omega t) \qquad 3$$

$$\lambda = \sqrt[4]{\frac{\rho A \omega^2}{EI}} \qquad \qquad 4$$

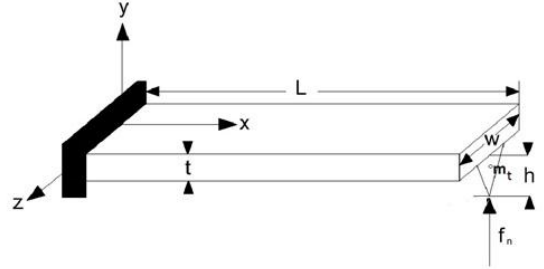

Fig. 1: normal forces exerted at the free end of micro-cantilever, the mass of the tip is considered as a lumped mass.

Where $a_1, a_2, ..., a_4$ are unknown constants and must be obtained from boundary conditions. We introduce constants A, B, C, D and $\Omega$ as follow:

$$A = \cosh\Omega + \cos\Omega$$
$$B = \sinh\Omega + \sin\Omega$$
$$C = \cosh\Omega - \cos\Omega \qquad \qquad 5$$
$$D = \sinh\Omega - \sin\Omega$$
$$\Omega = \lambda L$$

Using first three boundary conditions with respect to Eq.5 will gives:

$$y(x,t) = a_1 \{[\cosh(\lambda x) - \cos(\lambda x)] - \frac{A}{B}[\sinh(\lambda x) - \sin(\lambda x)]\} \sin(\omega t) \qquad 6$$

Using Eq.3 into the last boundary condition we have:

$$EI a_1 \lambda^3 (D - \frac{A^2}{B})\sin(\omega t) = k_1 a_1 (C - \frac{AD}{B})\sin(\omega t) +$$
$$k_n a_1^3 (C - \frac{AD}{B})^3 \sin^3(\omega t) - ... \qquad 7$$
$$m_t \lambda^2 C^2 a_1 (C - \frac{AD}{B})\sin(\omega t) + F\sin(\omega t + \phi)$$

$\Phi$ is the phase difference between cantilever and sample vibration and F is the external forces exists because of sample vibration and in the case of zero phase difference between sample and cantilever vibration is equal to:

$$F = \varepsilon a_n \{k_1 + k_n a_1 (C - \frac{AD}{B})^3\} \qquad 8$$

Let us make Eq.7 dimensionless. In this matter we introduce the dimensionless Amplitude of free end of micro-beam:

$$Y_a = y(L)/L = a_d(C - \frac{AD}{B})$$

$$a_d = \frac{a_1}{L} \qquad 9$$

Which $C_d$ is a dimensionless constant. By rewriting the Eq.7 in dimensionless form and using the dimensionless amplitude of free end of micro-beam, after some calculations we have:

$$\{\Omega^3(\frac{BD-A^2}{BC-AD})Y_a - \alpha Y_a + \mu\Omega^4 Y_a - \frac{3}{4}\beta Y_a^3 + P\}\sin\omega t + \{\frac{1}{4}\beta Y_a^3\}\sin 3\omega t = 0 \qquad 10$$

In which $\alpha = K_1 L^3 / EI$ is the dimensionless linear spring representing the linear interactions, $\beta = K_n L^5 / EI$ nonlinear dimensionless spring representing the nonlinear part of interactions, $\mu = \frac{m}{M}$ is tip-mass ratio (the ratio of probe mass to the cantilever mass) and $P = FL^2/EI$ is dimensionless effective force exerted to the cantilever in the case of both cantilever and sample vibrate in the same phase.

To satisfy Eq.10 at all times, both parts of Eq.10 must be zero simultaneously. This could happen just when the nonlinearity become zero ($\beta = 0$). In the nonlinear case, an approximation can be done when the first part of Equation abolish. The second part of Eq.10 will show that how much the approximation is close to exact value. The coefficient in second part of Eq.10 ($\frac{1}{4}\beta Y_a^3$) shows that only in vicinity of zero values of the micro-beam vibration amplitude of the free end of micro-beam this assumption can be true. In this case:

$$\Omega^3(\frac{BD-A^2}{BC-AD})Y_a - \alpha Y_a + \mu\Omega^4 Y_a - (\frac{3}{4})\beta Y_a^3 + P = 0 \qquad 11$$

Solving Eq.11 the Frequency response function of the cantilever will be revealed. In the linear case and without external forces, the natural frequency of the micro-beam will be determined approximately:

$$\Omega^3(\frac{BD-A^2}{BC-AD})Y_a - \alpha Y_a + \mu\Omega^4 Y_a = 0 \qquad 12$$

And in the nonlinear case, the Amplitude of vibration, without considering the sample vibration, is:

$$Y_a = \sqrt{\frac{4}{3\beta}[\Omega^3(\frac{BD-A^2}{BC-AD}) + \Omega^4\mu - \alpha]} \qquad 13$$

As we discussed before, Eq.10 does not satisfy exactly. This is because the Eq.6 indicates that required solution must contain higher orders of harmonics. This is why we introduce the Super harmonic Response in next chapter.

### 2.2. *Super Harmonic Response*

As it was mentioned in previous chapter, a response which satisfies the boundary conditions and equation of the motion must contain higher orders of harmonics. So we assume that the response of micro-beam as:

$$y(x,t) = \sum_{n=1,3,5,...} Y_n(x)\sin r\omega t \qquad 14$$

And the response of the micro-beam will be of the form:

$$y_r(x) = a_{1r}\cosh(\lambda x r^{1/2}) + a_{2r}\cos(\lambda x r^{1/2}) + a_{3r}\sinh(\lambda x r^{1/2}) + a_{4r}\sin(\lambda x r^{1/2}) \qquad 15$$

Again using first three boundary conditions will yield:

$$y(x,t) = a_1[\cosh(\lambda x) - \cos(\lambda x) - \frac{A_1}{B_1}(\sinh(\lambda x) - \sin(\lambda x))]\sin\omega t + a_3[\cosh(\sqrt{3}\lambda x) - \cos(\sqrt{3}\lambda x)... - \frac{A_3}{B_3}(\sinh(\sqrt{3}\lambda x) - \sin(\sqrt{3}\lambda x))]\sin 3\omega t \qquad 16$$

Next we introduce the dimensionless first harmonic and second harmonic amplitudes:

$$Y_{a1} = \frac{a_1}{L}(\frac{C_1 - A_1 D_1}{B_1})$$

$$Y_{a3} = \frac{a_3}{L}(\frac{C_3 - A_3 D_3}{B_3})$$

Solving the last boundary condition (Eq.2.e) and taking $A_n = \cosh(n\Omega) + \cos(n\Omega)$, $B_n = \sinh(n\Omega) + \sin(n\Omega)$, $C_n = \cosh(n\Omega) - \cos(n\Omega)$, and $D_n = \sinh(n\Omega) - \sin(n\Omega)$ will yield:

$$Y_{a1}^3 - Y_{a3}Y_{a1}^3 + \{\frac{4}{3\beta}[\alpha - \Omega^3(\frac{B_1D_1 - A_1^2}{B_1C_1 - A_1D_1}) - \Omega^4\mu]$$
$$+ 2Y_{a3}^2\}Y_a - \frac{4P}{3\beta} = 0$$

$$Y_{a3}^3 + \{\frac{4}{3\beta}[\alpha - 3\sqrt{3}\Omega^3(\frac{B_3D_3 - A_3^2}{B_3C_3 - A_3D_3}) - 9\Omega^4\mu]$$
$$+ 2Y_{a1}^2\} - \frac{1}{3}Y_{a1}^3 = 0$$

18

By solving the Eq.18, the response function of the micro-beam including the first and third order harmonics will be determined.

### 3. Case study and discussion

Comparison between the linear and nonlinear vibration of AFM reveals the sensitivity of resonant frequency to some parameters of the cantilever in the nonlinear interaction forces. In some parts of the curve, there is more than one response for the amplitude of vibration. Fig.2 shows the obtained frequency response curve in the case of considering/neglecting the mass of the tip in the linear and nonlinear case. The obtained FRF (Frequency Response Function) shows that in the linear vibration, FRF and also resonant frequency is sensitive to the tip-mass ratio (Ratio of the tip-mass to the cantilever mass), especially in the higher resonant frequency. The tip-mass ratio changed from zero ($\mu = 0$) to 7% of total mass of the cantilever ($\mu = .07$) according to the previews researches [8]. In this case, obtained resonant frequency by considering the tip-mass decreased by 12% (fig.2.a) while in the nonlinear case, the resonant frequency hasn't prominent change. The obtained FRF in the nonlinear case shows the least sensitivity to the tip-mass, especially in the primary resonant frequency. We considered the tip-mass to change from zero to 7% of the total mass in the nonlinear case. Although in the linear case the resonant frequency greatly affected by the tip-mass, in the nonlinear case, the deviation from the resonant frequency is merely negligible (Fig.2.b). While in reality the exerted interaction forces are nonlinear, so it is expected to see less effect in resonant frequency due to the tip-mass change, especially for the commercial kinds of AFM.

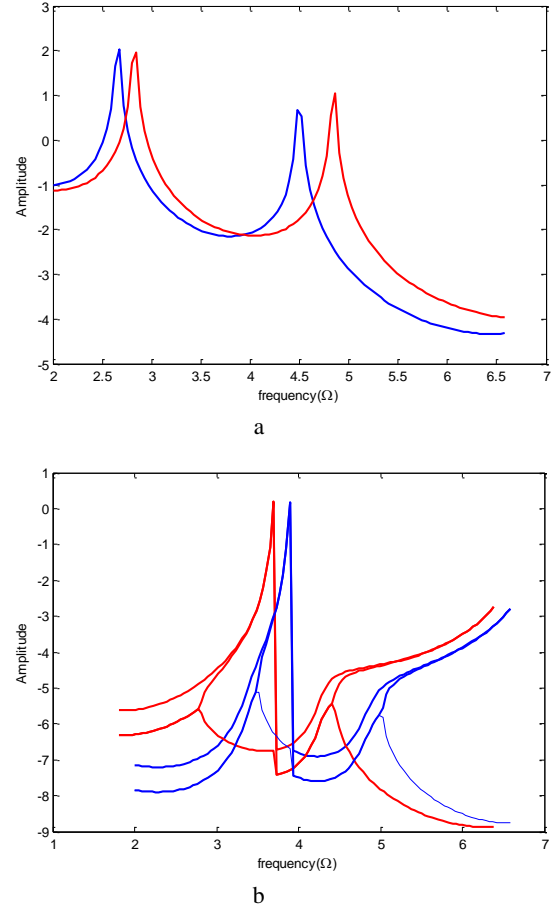

Fig. 2: Effect of tip-mass is obtained. The red line shows the FRF of the cantilever when the mass of the probe is neglected and the blue line shows the case when the mass of the tip is 7% of the mass of cantilever. In the linear case, by increasing the tip-mass the resonant frequency decrease, while in the nonlinear interaction assumption, the resonant frequency shows no sensitivity to the probe mass.

Fig.3 and Fig.4 show the effect of sample vibration on the nonlinear dynamics of AFM. The frequency response curve is obtained for different values of epsilon ($\varepsilon$). In fact, the vibration amplitude of sample can be determined by $\varepsilon$. The FRF curve is obtained for different values of epsilon parameter. When there is no phase differences between sample and cantilever vibration, the vibration of the sample causes the both resonant frequency and the amplitude of the micro-beam changes extravagantly. As the vibration amplitude of sample increases, the resonant frequency of cantilever increases considerably (Fig.3). This is because the interaction forces decrease due to the in-phase vibration of cantilever and sample. Even if the sample vibrates at the very small amplitude compared to the vibration of cantilever, still it has the considerable effect on the dynamics of AFM. The FRF of the cantilever when the sample vibrates at the anti phase amplitude is studied too. It is obtained that the situation reverses at anti-phase vibration of the sample. By increasing the amplitude of sample, the resonant frequency of cantilever decreases. This is because the distance between tip and sample increases and the interaction forces has less effect on the dynamics of cantilever. Fig.4 shows the results; the obtained resonant frequency response is obtained for different values of $\varepsilon$.

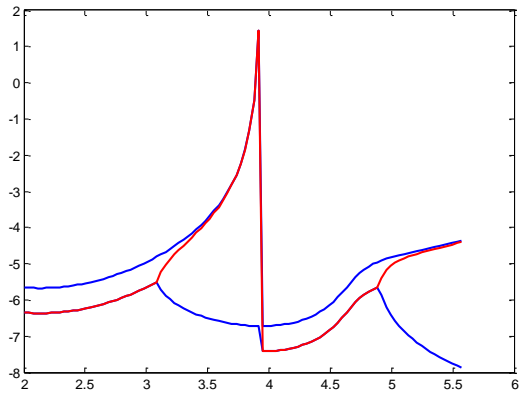

Fig. 3: The effect of in-phase vibration of sample on resonant frequency of AFM is obtained. Red line shows the FRF when $\varepsilon = 0$ and the blue line shows the FRF of the cantilever when $\varepsilon = 0.1$. By increasing the amplitude of the sample, the resonant frequency of cantilever increases.

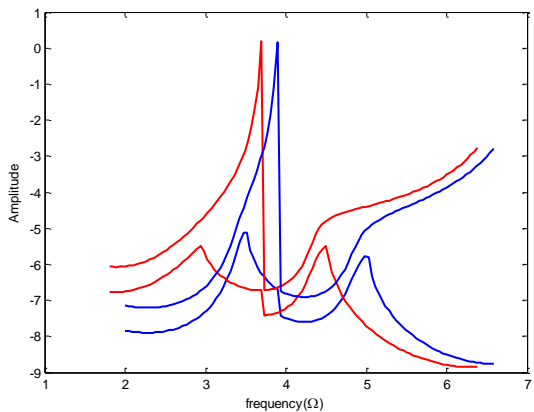

Fig. 4: The effect of anti-phase vibration of the sample on the FRF of the cantilever is obtained. The blue line shows the FRF when $\varepsilon = 0$ and the red line shows the FRF of the cantilever when $\varepsilon = 0.1$. By increasing the amplitude of the sample vibration, the resonant frequency decreases.

## 4. Conclusion

The analytical method is developed in order to study the dynamics of AFM in the case of nonlinear interaction forces. The exerted interaction forces in the nonlinear case are simulated with a polynomial with the minimum error. The resonant frequency curve is obtained in the linear and nonlinear case taking into account the sample vibration. It is obtained that although the resonant frequency is sensitive to the tip-mass ration in the linear case, it is not sensitive to the tip-mass ratio in the nonlinear case, especially in the primary resonant frequency. The obtained resonant frequency curve for different values of tip-mass ratio testifies this. The effect of in-phase and anti-phase vibration of the sample is studied. It is obtained that the resonant frequency increases when the vibrating amplitude of sample increases in the in-phase vibration. The situation reverses when the sample vibrates at anti-phase vibration. This means that resonant frequency decreases when the vibration amplitude of micro-beam increases.